\documentclass{quantumview}
\usepackage[utf8]{inputenc}
\usepackage[T1]{fontenc}
\usepackage{rotating} 
\usepackage{graphicx,bm,bbm}
\usepackage{amsmath,amssymb}
\usepackage{hyperref}
\usepackage[english]{babel}
\usepackage[numbers,sort&compress]{natbib}
\usepackage{caption}
\begin{document}

\title{Avoiding Defects within Heavy-Hexagon Architecture}
\author{Hansol Kim}
\email{khshk18@hanyang.ac.kr}
\author{Younghun Kwon}
\email{yyhkwon@hanyang.ac.kr}
\affiliation{Department of Applied Physics, Hanyang University (ERICA), Ansan 15588, Republic of Korea}
\begin{abstract} 
\indent When implementing scalable fault-tolerant quantum computing using superconducting quantum processors, one of the critical challenges to overcome is the presence of defective qubits arising from fabrication and calibration imperfections. A defect on a specific qubit severely restricts the connectivity between adjacent qubits, disrupts stabilizer measurements, induces frequent detection events, and consequently inflates the logical error rate. To mitigate this issue, various strategies have been proposed, such as syndrome extraction circuits based on stabilizer folding and the construction of super-stabilizers via gauge operators. However, these methods have primarily been developed for architectures with relatively high connectivity of 3 to 4. Furthermore, experimental demonstrations showing an actual reduction in logical error rates in the presence of physical hardware defects have been lacking. In this work, we propose a defect-avoidance scheme tailored for the heavy-hexagon architecture, which features a lower connectivity of 2 to 3. While previous approaches required four measurement cycles to measure all gauge operators comprising a super-stabilizer, our proposed scheme completes all gauge measurements within only two cycles. We experimentally validate this approach on IBM quantum hardware by identifying physical defects and implementing the defect-avoidance circuits. As a result, avoiding the defect yields an approximately 10\% reduction in the logical error rate per cycle for both logical $X$ and $Z$ errors. These findings demonstrate that defects in superconducting hardware can be successfully handled even under the constrained connectivity of heavy-hexagon lattices, and that defect avoidance substantially reduces logical errors on real quantum hardware.
\end{abstract} 

\maketitle
%
%

\section{Introduction}
\indent Fault-tolerant quantum computation is essential for unlocking the transformative potential of quantum algorithms~\cite{Deutsch, Simon, Shor, Grover, Vazirani, Harrow, Campbell, kivlichan2020improved}. To protect fragile quantum states from decoherence and environmental noise, quantum error correction (QEC) protocols encode logical information non-locally across multiple physical qubits, founded upon the stabilizer formalism and topological code architectures~\cite{stabilizer, Daniel, boundary, topological, anyons, surface, Fowler2, qecmemory, Roffe}. With rapid advancements in quantum hardware technologies, quantum error correction, which was once discussed primarily in theoretical terms, is now being realized experimentally across a diverse range of physical platforms, including superconducting circuits~\cite{ibm_bell, magic, kim2025effectivenesssyndromeextractioncircuit, google2025quantum, eickbusch2024demonstrating, d2_qec, d3_surf, d3_surper, ibm_rep, google_qec2}, neutral-atom arrays~\cite{neutral, Evered, nutral_atom}, and trapped ions~\cite{d3_bacon, d3_ion, Moses}. Nevertheless, most demonstrations to date have been confined to small scales, with the largest experimental implementation achieving a distance-7 surface code~\cite{google2025quantum}. A prominent hurdle hindering further scaling is the presence of physical defects in superconducting chips. Owing to fabrication imperfections, frequency crowding, or calibration instabilities, specific qubits or couplers on a two-dimensional lattice can exhibit unacceptably high physical error rates, rendering them unusable~\cite{hardware, Zhang, morvan2022optimizing, ecr, virtual_Z, rd}. Such defects trigger high detection event rates and contribute substantially to the logical error of the code. Furthermore, as the hardware footprint expands, the probability of encountering defects increases, presenting severe obstacles to scaling up QEC codes~\cite{debroy2025luci, wolanski2025automated, drouet2026error, wei2026adaptive, strikis2023quantum, siegel2023adaptive, lin2024codesign, kim2026luci}.

To address this defect problem, several strategies have been proposed that modify the stabilizers of the surface code to avoid defective elements. These approaches enable effective defect avoidance either with a minor reduction in code distance or with no distance reduction at all. However, they generally require a connectivity of degree~4, in which each qubit is coupled to four nearest neighbors~\cite{PhysRevA.96.042316, strikis2023quantum, siegel2023adaptive, lin2024codesign, leroux2025snakes, mishmash2025excising, zhou2024halma, min2026defectadaptive, anker2025optimized, rathi2026xz}. 

In recent years, novel syndrome measurement schemes, such as stabilizer folding, have stimulated extensive research into implementing QEC codes on architectures with lower degree connectivity, including hexagonal codes on degree-3 hexagonal lattices and surface codes on heavy-hexagon lattices~\cite{flag4, flag3, Almudever, ibm_qec2, hh-code, kim2025implementation, Wu, flag, flag2}. In hardware with such reduced connectivity, distinct methodologies are demanded for defect handling. Recently, a defect-avoidance scheme for hexagonal codes on degree-3 hexagonal lattices was proposed by building on previous work~\cite{higgott2025handling}. However, the heavy-hexagon architecture widely adopted by IBM quantum systems possesses an even lower connectivity of $2\sim 3$~\cite{abughanem2024ibm}. On heavy-hexagon lattices, prior studies have adapted degree-3 hexagonal codes by re-introducing syndrome qubits that had been eliminated in the hexagonal code formulation and employing additional ancilla qubits~\cite{benito2025comparative, vezvaee2025surface, kim2026magic, math13233874, zen2026lowvalency, lee2026scalable}.

In this work, we optimize and construct scalable syndrome extraction circuits tailored to real heavy-hexagon hardware, enabling the realization of distance-7 surface codes on current IBM quantum systems. Furthermore, we extend defect-avoidance techniques from hexagonal lattices to the heavy-hexagon lattice, demonstrating that defects can be effectively handled even in systems with such low connectivity. By executing these protocols on physical quantum processors, we confirm that reconfiguring stabilizers to avoid actual hardware defects yields a measurable reduction in the logical error rate.

\section{Method}

\subsection{Rotated surface code and mid-circuit state}
\begin{figure}[ht]

\centerline{\includegraphics[width=1\linewidth]{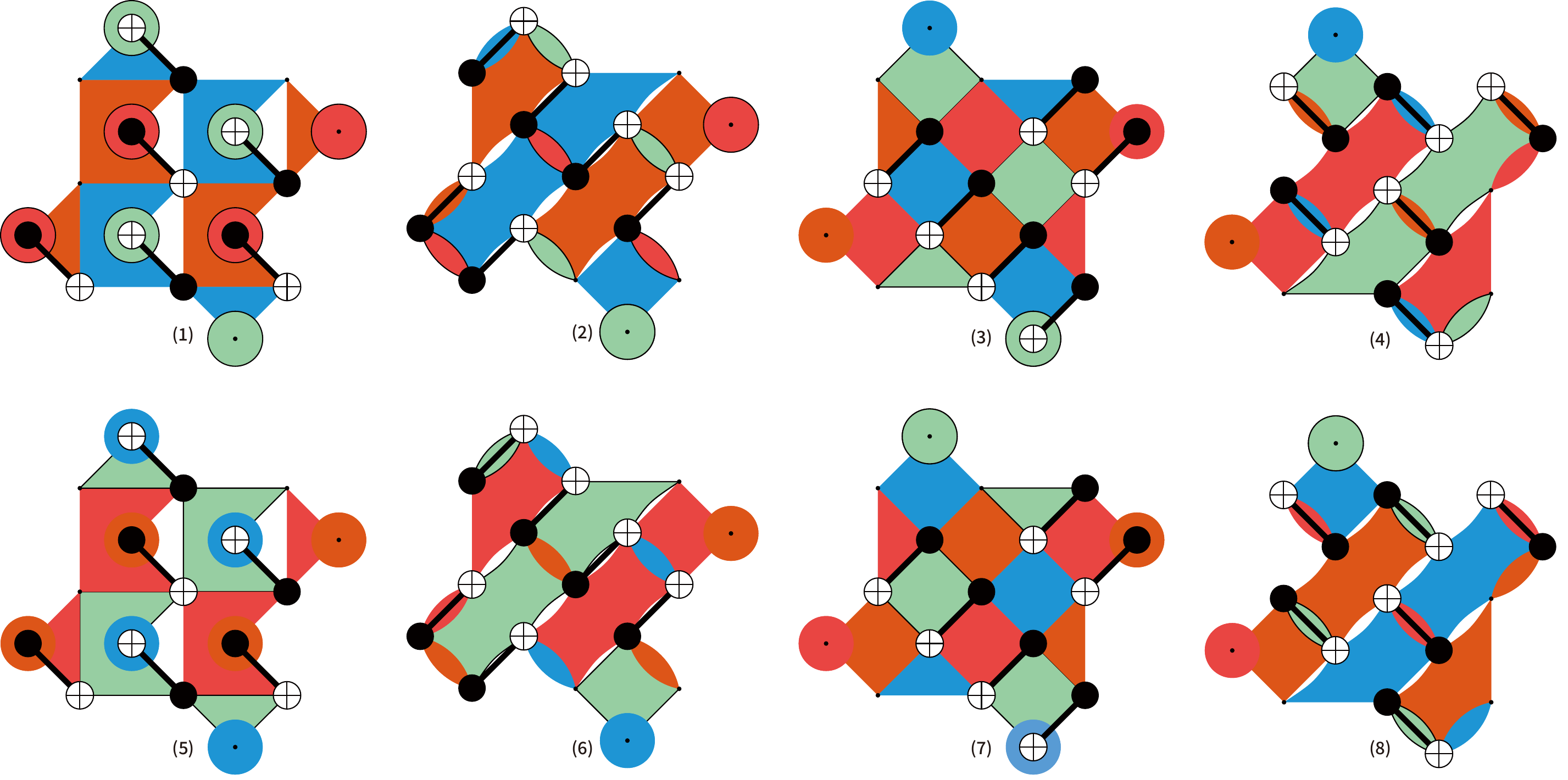}}

\caption{Mid-circuit states of the rotated surface code. Pauli $X$ stabilizers are shown in red and orange, while Pauli $Z$ stabilizers are shown in blue and green. During the application of CNOT gates for stabilizer measurements, the stabilizer operators transform, and a standard surface code emerges as an intermediate mid-circuit state. After one measurement round, the red and orange stabilizers exchange roles, as do the blue and green stabilizers. Across two consecutive rounds, states (1) through (8) undergo a periodic cycle.}

\label{fig:rotated surface code}

\end{figure}
\begin{figure}[ht]

\centerline{\includegraphics[width=0.7\linewidth]{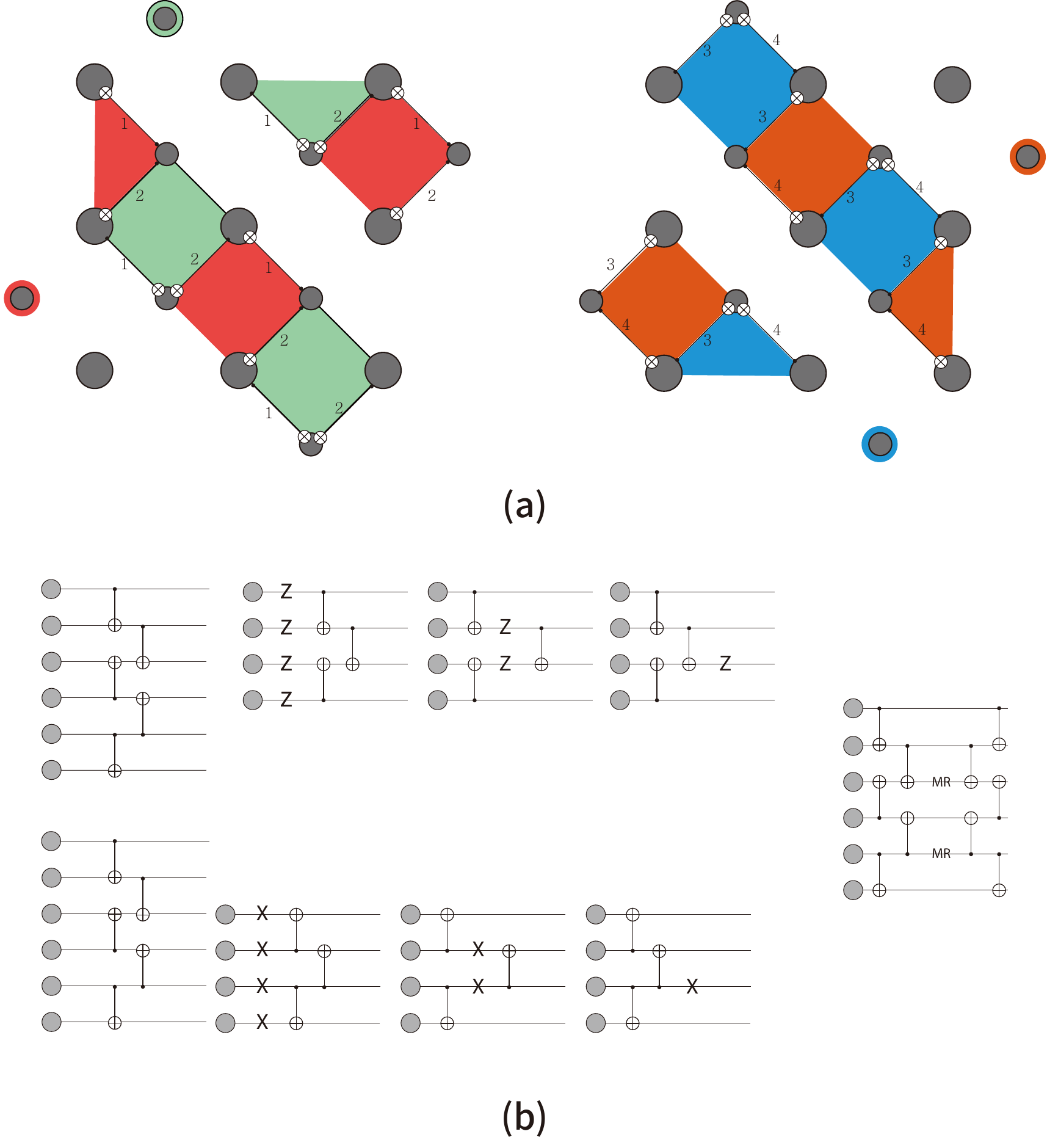}}

\caption{(a) Schematic representation of dividing the mid-circuit state (7) from Fig.~\ref{fig:rotated surface code} in half and applying CNOT gates. Each stabilizer is folded through two CNOT layers into a localized stabilizer that can be measured on a single qubit. (b) Quantum circuit diagram illustrating the folding process in (a). By folding the stabilizer onto a single qubit for measurement and subsequently unfolding it, the system returns to its original state.}

\label{fig:stabilizer folding}

\end{figure}
The rotated surface code is an error-correcting code that rotates the geometry of the standard surface code by $45^\circ$ and modifies the boundary stabilizers, achieving the same code distance with fewer physical qubits~\cite{surface, Fowler2, Bravyi2}. Because the bulk stabilizer layout is a rotated version of the surface code, it consists of weight-4 stabilizers composed of Pauli $X$ and $Z$ operators:
$$
S_Z^4 = \langle Z_a Z_b Z_c Z_d \rangle, \quad S_X^4 = \langle X_a X_b X_c X_d \rangle
$$
The rotated surface code is constructed by arranging these two types of stabilizers periodically across a planar lattice. However, constructing the code exclusively from weight-4 stabilizers leaves errors on the boundary data qubits undetectable. Therefore, weight-2 stabilizers commuting with all surrounding weight-4 stabilizers are incorporated along the boundaries to identify distinct error configurations:
$$
S_Z^2 = \langle Z_a Z_b \rangle, \quad S_X^2 = \langle X_a X_b \rangle
$$
Under this configuration, operators that commute with all stabilizers serve as logical operators (or logical errors), providing the mathematical foundation for fault-tolerant logical operations~\cite{FT1, clifford1, clifford2, Eastin, Zeng}.

To implement the rotated surface code on physical quantum processors, stabilizer measurements must be realized via quantum gate sequences. Circuits for stabilizer measurement in rotated surface codes typically introduce syndrome qubits positioned among data qubits to execute CNOT (or CZ) gate scheduling, and the optimal gate scheduling orders have been extensively studied. When accounting for circuit-level errors such as hook errors, it has long been established that scheduling CNOT (CZ) gates in `N' and `Z' orders for $X$ and $Z$ stabilizers provides efficient fault tolerance~\cite{kishony2026surface}. Recent studies have analyzed the lattice representation of these circuits, identifying that standard surface codes arise as intermediate mid-circuit states during rotated surface code execution~\cite{mcewen2023relaxing} (Fig.~\ref{fig:rotated surface code}). In this mid-circuit standard surface code, the qubits conventionally assigned as syndrome ancillas are treated on equal footing with data qubits, regarding all physical qubits as data qubits. By orchestrating the CNOT gate schedule that governs transitions between the mid-circuit standard surface code and the rotated surface code, a stabilizer measurement circuit realizable with degree-3 connectivity without dedicated syndrome ancillas was proposed (Fig.~\ref{fig:stabilizer folding}). This architecture folds each stabilizer onto a single data qubit, measures that qubit, and unfolds it back to the original state---a technique termed stabilizer folding. This paradigm enables two key variants of the surface code: first, implementation on degree-3 hexagonal lattices, and second, leveraging spare connectivity to bypass defective couplers~\cite{debroy2025luci}.

\subsection{Round and Cycle}
In the quantum error correction literature, the terms `cycle' and `round' are frequently used to quantify time intervals. In conventional surface code experiments, these two terms are often used interchangeably. However, in recent syndrome extraction schemes such as stabilizer folding, not all stabilizer operators are measured within a single measurement cycle. To eliminate terminological ambiguity, we define `round' and `cycle' throughout this paper as follows: a \emph{round} is the time unit during which every stabilizer operator defining the code is measured at least once; a \emph{cycle} is the time unit during which a designated subset of stabilizers is measured simultaneously. When all stabilizers can be measured concurrently---as in standard surface codes or conventional rotated surface codes---round and cycle become equivalent. Conversely, when stabilizer measurements are partitioned across multiple stages, these concepts become distinct. Furthermore, we subdivide each round into cycle units referred to as \emph{subrounds}. Through this nomenclature, individual cycles are denoted as subround~1, subround~2, etc., explicitly indicating that different subsets of stabilizer operators are being measured in each stage.

\subsection{Hexagonal Code and Implementation on Heavy-Hexagon Architecture}
\begin{figure}[ht]

\centerline{\includegraphics[width=1\linewidth]{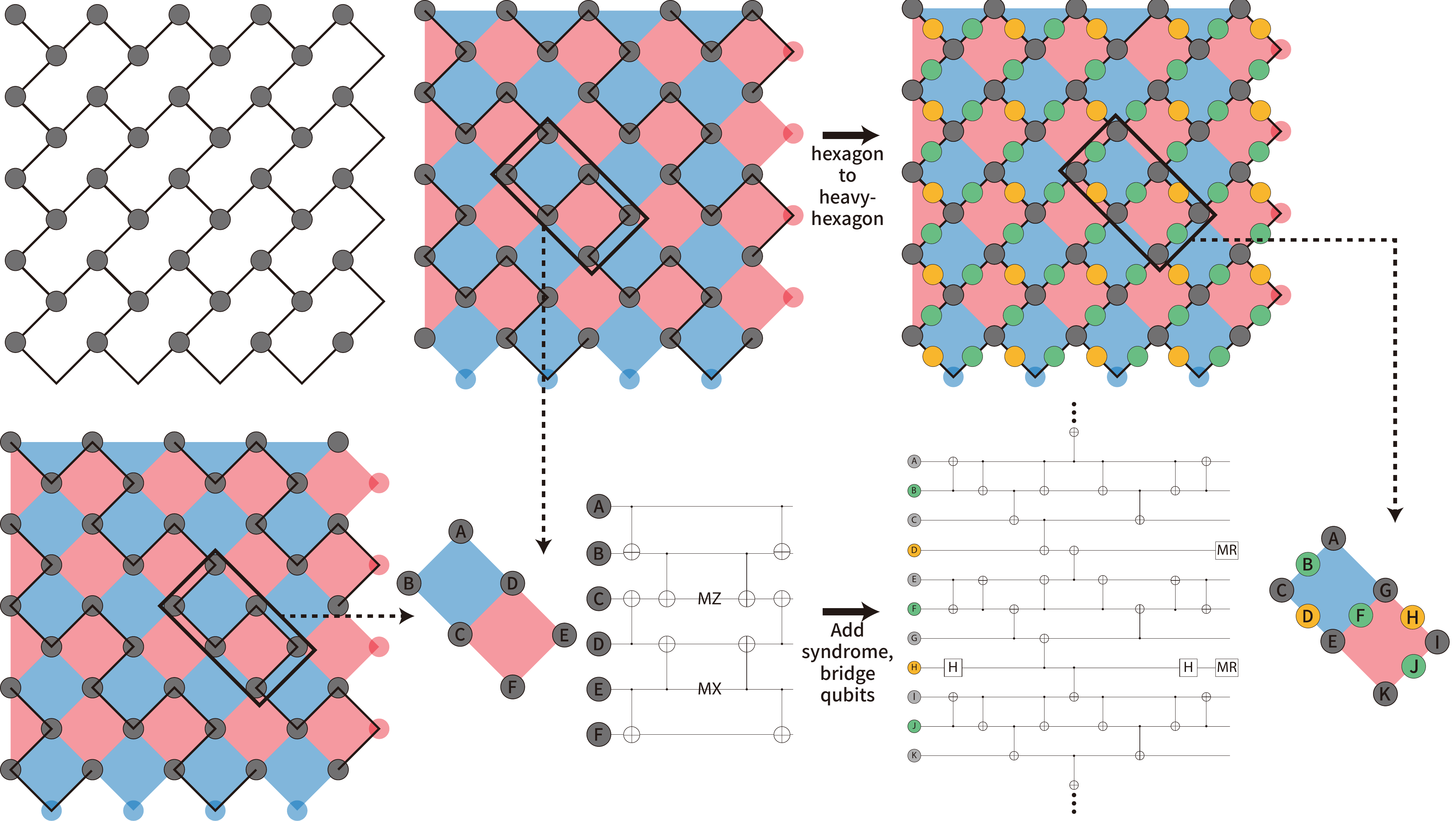}}

\caption{Hexagonal code realizing a standard surface code on a degree-3 hexagonal lattice. All stabilizer measurements are completed across two measurement cycles. In each cycle, a column of qubits is selected, and two layers of CNOT gates are applied among them to fold the weight-4 stabilizers of that array onto single qubits, thereby forming the syndrome extraction circuit. Following measurement, the folded state is unfolded to restore the original state before initiating the next cycle. This architecture can be mapped onto the heavy-hexagon layout, where syndrome qubits (yellow circles) are naturally re-introduced. This re-introduction allows constructing syndrome measurement circuits without an explicit unfolding phase. Due to the limited connectivity of the heavy-hexagon architecture, incorporating SWAP sequences via additional bridge qubits (green circles) enables the implementation of standard surface codes analogously to the hexagonal code.}

\label{fig:hexagonal_code}

\end{figure}
By employing stabilizer folding inspired by the mid-circuit state of the rotated surface code, weight-4 stabilizers of the surface code can be realized on a degree-3 hexagonal lattice, yielding what is known as the hexagonal code. While the hexagonal code shares the same fundamental code space and stabilizer group as the standard surface code, the limited connectivity along the right and lower boundaries necessitates adding extra qubits with associated weight-1 stabilizers. In the hexagonal code, all stabilizers cannot be measured concurrently in a single step; hence, each round is divided into two subrounds, measuring all stabilizers across two cycles (Fig.~\ref{fig:hexagonal_code}).

Previous research explored applying this hexagonal code concept to the heavy-hexagon architecture~\cite{benito2025comparative, vezvaee2025surface, zen2026lowvalency, lee2026scalable}. Because the heavy-hexagon lattice contains additional qubits placed along the edges of the hexagonal tiles, a portion of these additional qubits can naturally serve as syndrome ancillas. Re-introducing syndrome qubits eliminates the need for the unfolding process used in hexagonal codes to reset the state after measurement. However, the auxiliary qubits that do not serve as syndrome ancillas restrict available direct routing, requiring additional CNOT gates to implement SWAP operations. A notable feature of this architecture is that syndrome qubits are shared between two distinct types of stabilizers. This sharing causes the creation and destruction dynamics of stabilizers to diverge from standard surface codes. While this strategy utilizes heavy-hexagon qubits to realize surface codes, it is fundamentally optimized for rotated planar geometries, leading to hardware inefficiencies when deployed on physical IBM devices.

\subsection{Stabilizer Measurement on Heavy-Hexagon Architecture}
\begin{figure}[!t]
\centerline{\includegraphics[width=1\linewidth]{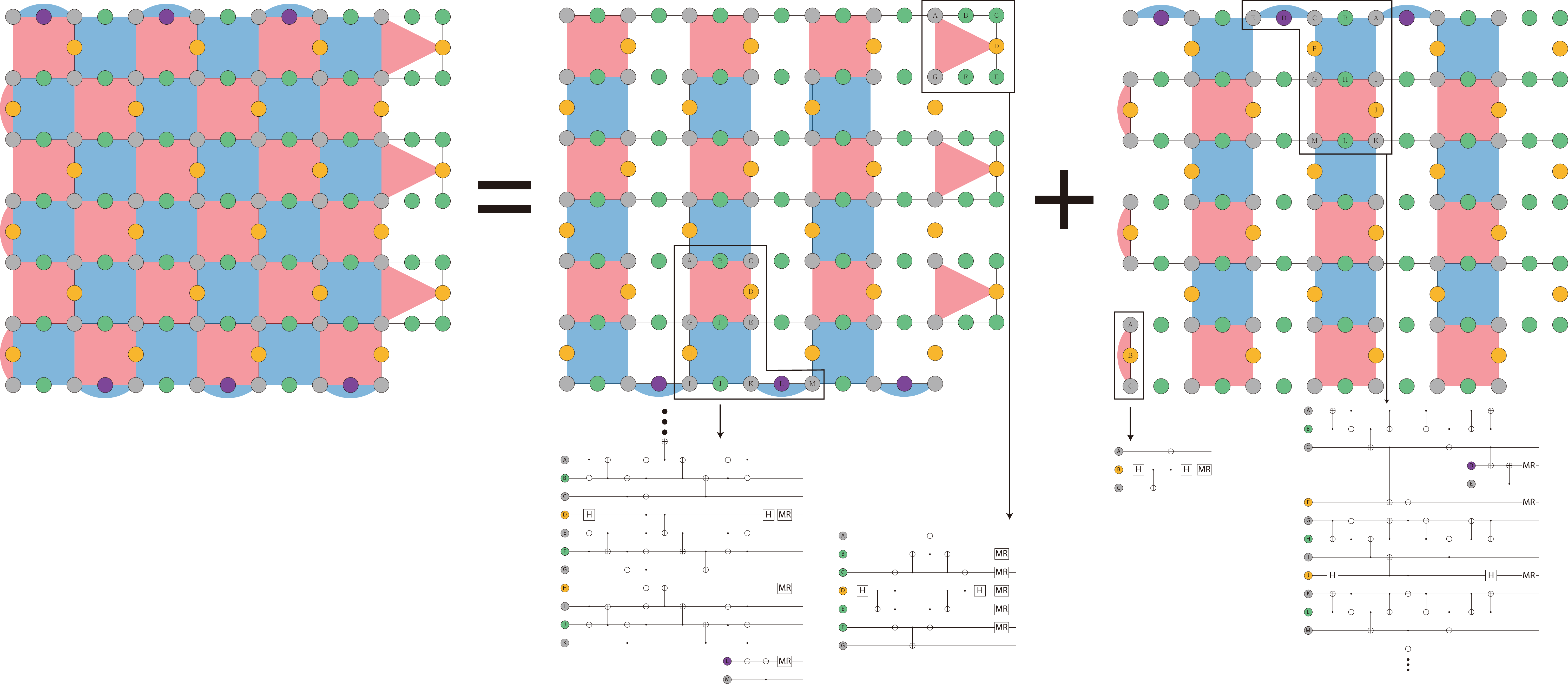}}
\caption{Structure of the rotated surface code implemented on the heavy-hexagon architecture. Weight-4 stabilizers are realized via stabilizer folding, while weight-2 boundary stabilizers along the top and bottom edges are appended to the final stage of the stabilizer folding sequence. Stabilizers along the right boundary are measured in the first subround using two flag qubits per data qubit, whereas stabilizers along the left boundary are measured directly without flag qubits in the second subround.}
\label{fig:surface code on HH with flag}
\end{figure}

To implement weight-4 stabilizer measurements of the rotated surface code on the heavy-hexagon architecture, we fundamentally employ the stabilizer folding method. However, this approach is impractical for measuring the weight-2 boundary stabilizers. Therefore, we adapted syndrome extraction circuits developed in other architectures to measure the weight-2 boundary stabilizers. On the heavy-hexagon lattice, the connectivity along the top, bottom, and left boundaries naturally accommodates syndrome qubits for weight-2 stabilizer extraction. Along the right boundary, however, direct connectivity between adjacent data qubits is absent. To bridge this connectivity gap, we utilize flag qubits to mediate indirect coupling between data qubits and syndrome ancillas~\cite{flag, flag2, flag3, Almudever, hh-code}. As illustrated in Fig.~\ref{fig:surface code on HH with flag}, we configure stabilizer measurement circuits that measure data qubits in the Pauli $X$ basis using two flag qubits per data qubit. By applying a Hadamard gate to the syndrome qubit and executing a sequence of CNOT gates across the flag qubits---with controls directed toward the syndrome qubit and targets toward the data qubits---errors on data qubits propagate faithfully to the syndrome ancilla. Along the top and bottom boundaries, measurements must be scheduled without conflicting with the ongoing stabilizer folding process. To achieve this, we introduce circuits measuring the upper and lower weight-2 stabilizers across two subrounds, which increases the circuit depth by the equivalent of one CNOT gate.

In this manner, we establish a scheme capable of executing all rotated surface code stabilizer measurements within two subrounds. Under this layout, rotated surface codes can be implemented on heavy-hexagon hardware using only 23 qubits for distance 3, 65 qubits for distance 5, and 127 qubits for distance 7. In general, for an odd code distance $d$, our construction requires $N = \frac{1}{2}(5d^2 + 2d - 5)$ physical qubits. Compared to previous architectures requiring $10d^2 - 8d - 1$ or $5d^2 - 2d - 2$ qubits, this constitutes a 2- to 4-fold reduction in physical qubit overhead. Furthermore, by introducing minor adjustments to previously released IBM layouts---such as IBM Osprey (433 qubits) or IBM Condor (1121 qubits)---such as shifting corner qubit assignments or modifying syndrome extraction paths, it becomes possible to realize distances 13 and 21, respectively.

Nevertheless, this design has certain trade-offs. The principal drawback is that the direction of stabilizer folding aligns parallel to the logical operator. During stabilizer folding, numerous two-qubit gates are executed along the horizontal axis, and certain fault mechanisms during this sequence can propagate simultaneously to two data qubits within a single round (see Supplementary Information~\ref{supp:error_propagation}). For errors corresponding to the horizontal logical operator, such a fault behaves as an effective weight-2 error, reducing the effective code distance to $\lceil d/2 \rceil$ (i.e., $\lceil d/2 \rceil$ physical errors can induce a logical failure). In our experimental implementation, the logical $X$ operator lies along the horizontal direction; consequently, the logical $X$ error rate is amplified compared to the logical $Z$ error rate. Despite this reduction in effective distance, exponential error suppression remains attainable as long as physical error rates remain below the fault-tolerance threshold for logical $X$ errors, enabling arbitrary target error rates to be achieved by scaling code distance (see Supplementary Information~\ref{supp:threshold}). Expanding the heavy-hexagon aspect ratio preferentially along the horizontal direction could mitigate this asymmetric logical vulnerability. Moreover, because the defect-avoidance scheme proposed in this work applies equally to rotated lattices implementing standard surface codes, we adopt this architecture to maximize physical qubit utilization on current hardware.

\subsection{Prior Approaches to Defect Handling}
Physical defects introduced during hardware fabrication pose a substantial barrier to fault-tolerant QEC~\cite{hardware, Zhang, morvan2022optimizing, ecr, virtual_Z, rd}. In transmon-based superconducting quantum processors in particular, fabrication flaws or qubit frequency crowding during calibration can severely degrade individual qubit coherence or gate fidelity, leading to localized regions of high error rates. The limited connectivity of superconducting chips compounds this vulnerability. Hardware defects are broadly categorized into coupler defects (affecting two-qubit gates) and qubit defects (affecting single-qubit gates or readout). In conventional surface codes, qubit defects have historically been treated differently depending on whether the affected site is a data qubit or a syndrome qubit. Numerous studies have proposed merging neighboring stabilizers into `super-stabilizers' using gauge operators to bypass defects. This approach generally trades a small amount of code distance to form a super-stabilizer that absorbs the defective degree of freedom, effectively isolating the defect. The recently developed LUCI scheme demonstrated that coupler defects can be bypassed without distance loss by leveraging stabilizer folding, and constructed super-stabilizers for qubit defects~\cite{debroy2025luci}. However, that formulation relied on degree-4 connectivity. In degree-3 architectures such as hexagonal lattices, defect avoidance without distance reduction is not feasible; instead, Ref.~\cite{higgott2025handling} showed that merging adjacent non-defective stabilizers into gauge operators allows defects to be bypassed with a distance reduction of only 1, a scheduling principle further generalized in recent defect compilation frameworks~\cite{anker2025optimized, rathi2026xz}.

\subsection{Defect Handling on Heavy-Hexagon Architecture}
\begin{figure}[!t]
\centerline{\includegraphics[width=1\linewidth]{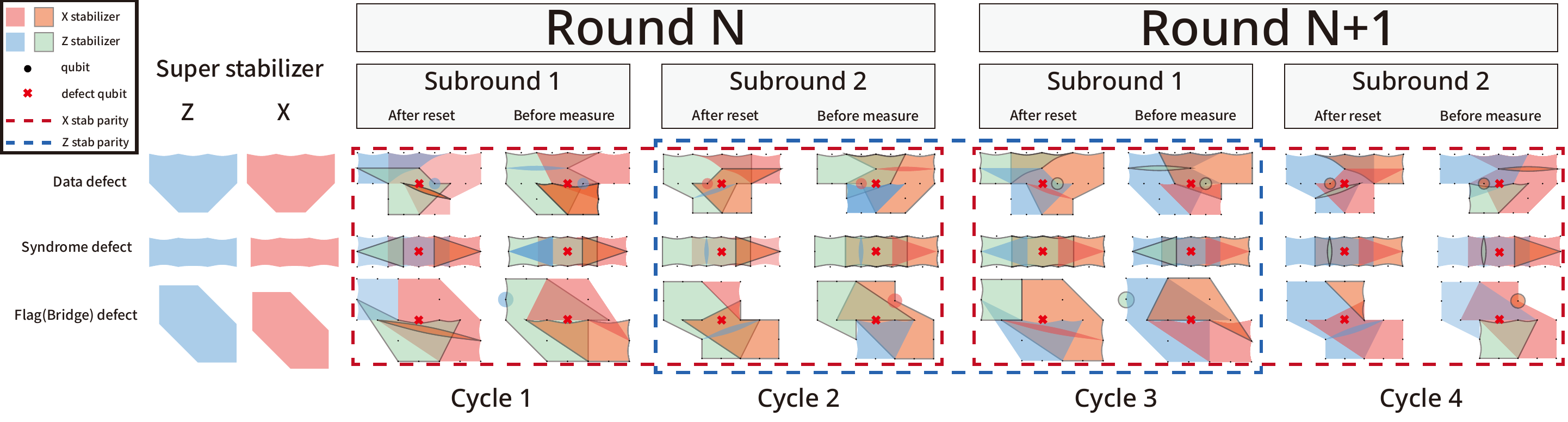}}
\caption{Construction of super-stabilizers for different defect types and the dynamical transformation of stabilizers on the heavy-hexagon architecture. To track stabilizer propagation mediated by syndrome ancillas, stabilizers are displayed in paired colors ($X$: red and orange; $Z$: blue and green). To complete the measurement of each super-stabilizer within two cycles, parity checks for $Z$ stabilizers are temporally staggered. Through this scheme, super-stabilizer eigenvalues are reconstructed from gauge measurements obtained across consecutive subrounds.}
\label{fig:Defect avoiding}
\end{figure}
We implement this defect-avoidance strategy on the rotated surface code mapped onto the heavy-hexagon architecture. While our basic super-stabilizer and gauge configurations build upon the hexagonal code methodology, a crucial distinction in the heavy-hexagon layout is the re-emergence of physical syndrome qubits, requiring careful consideration of the augmented stabilizer dynamics. In architectures lacking syndrome qubits, such as LUCI or the original hexagonal scheme, measuring all gauge operators comprising a super-stabilizer necessitated four measurement cycles. On heavy-hexagon hardware, however, the availability of dedicated syndrome ancillas and bridge (flag) qubits allows them to be repurposed as syndrome qubits for gauge extraction, completing all gauge measurements within just two cycles. Figure~\ref{fig:Defect avoiding} illustrates the configurations of super-stabilizers and gauge operators alongside the corresponding syndrome ancillas for different defect scenarios. To complete these measurements within two cycles, detectors must be constructed by tracking stabilizer evolution throughout the syndrome extraction sequence. We interleave the detector construction for $X$ and $Z$ super-stabilizers: for $X$ super-stabilizers, detectors are formed by comparing two subrounds within the same round, whereas for $Z$ super-stabilizers, detectors compare subround~2 of the preceding round with subround~1 of the current round. This interleaved schedule ensures that all constituent gauge operators are measured within two cycles, enabling robust super-stabilizer reconstruction. Detailed gate sequences and scheduling are provided in Supplementary Information~\ref{supp:gate_timeline}.

In the LUCI and hexagonal architectures, qubit defects and coupler defects lead to distinct gauge choices. In heavy-hexagon layouts, however, the strict connectivity constraints make the impact of coupler defects effectively identical to that of qubit defects. The heavy-hexagon lattice consists of degree-3 nodes connected exclusively to degree-2 nodes. Consequently, a coupler defect can be treated equivalently to a defect on an adjacent degree-2 qubit, allowing us to model physical errors primarily at the qubit level. Furthermore, defects on syndrome ancillas or flag (bridge) qubits---which were absent in pure hexagonal layouts but are analogous to coupler defect scenarios---require tailored super-stabilizer constructions, as depicted in Fig.~\ref{fig:Defect avoiding}. 

In real IBM calibration data, we frequently observe sites where coupler errors and qubit errors coincide. Such locations maintain persistently high physical error rates across repeated calibration cycles, pointing to permanent fabrication flaws. Focusing on these hardware imperfections, we implemented and experimentally tested our defect-avoidance protocol on physical quantum devices.

\section{Results}
\subsection{Defect selection}
We conducted experiments on \texttt{ibm\_boston}, a 156-qubit heavy-hexagon processor. IBM periodically recalibrates its quantum devices, causing gate error rates to fluctuate over time. However, when fabrication imperfections exist, certain qubits consistently exhibit high error rates or fail calibration altogether, regardless of tuning. On \texttt{ibm\_boston}, physical qubit~85 exhibited persistently degraded performance across all calibration cycles, accompanied by severely elevated error rates on all incident CZ gates. We identified this site as a permanent hardware defect and compared rotated surface codes incorporating the defect directly against codes configured with our defect-avoidance protocol. Because fabrication defects cannot be arbitrarily relocated on physical hardware, our experiments focused specifically on the data-qubit defect naturally present on \texttt{ibm\_boston}.

\subsection{Syndrome detection and correlation}
\begin{figure}[!t]
\centering
\includegraphics[width=1\linewidth]{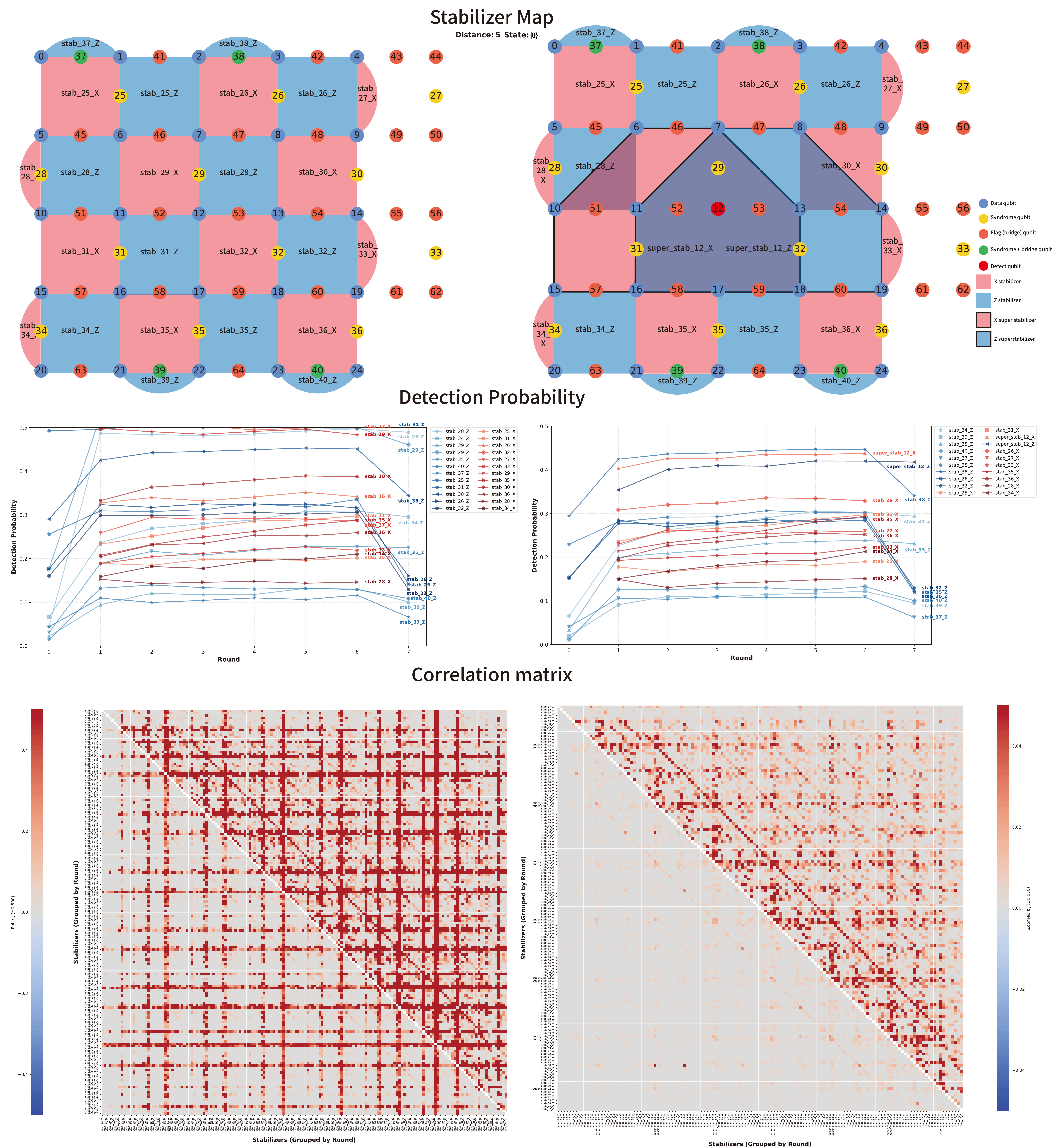} 
\caption{Detection event probabilities and correlation matrices for a distance-5 rotated surface code prepared in the $|0\rangle$ state, comparing configurations with and without defect avoidance. Stabilizers are labeled by the index of their measuring syndrome qubit and stabilizer type; super-stabilizers are indexed by the defective qubit. Around defective qubit~12 (corresponding to physical qubit~85), the detection probabilities of the four neighboring stabilizers approach 0.5, erasing syndrome information. When bypassing the defect with a super-stabilizer, the detection probability drops to approximately 0.4, representing an appreciable improvement. Furthermore, the defect-induced correlations caused by near-random detection events are significantly suppressed, revealing characteristic error signatures such as time-like fault propagation.}
\label{fig:Detection event and corrletaion}
\end{figure}

By analyzing syndrome strings extracted across measurement rounds, one can identify which stabilizer operators frequently flag error events. In the presence of a hardware defect, the detection probability of neighboring stabilizers approaches 0.5, completely erasing syndrome information. We analyzed experimental syndrome strings to extract detection probabilities across all stabilizer operators (Fig.~\ref{fig:Detection event and corrletaion}). Stabilizers coupled to the defective qubit show detection probabilities saturating near 0.5. Near-random measurement outcomes signify that syndrome measurements yield zero mutual information regarding physical faults, severely degrading decoding and error correction. By implementing our defect-avoidance protocol, however, the previously random detection probabilities are suppressed to approximately 0.4. Although gauge extraction requires additional readout operations---the dominant error-inducing primitive---leading to slightly higher detection rates for super-stabilizers compared to defect-free stabilizers, the detection probability is measurably lower than that of the unmitigated defective configuration.

Detection events also allow us to compute two-point error correlations. From individual detection probabilities, we evaluate how frequently detection events co-occur, calculating the correlation $p_{ij}$ between detectors $i$ and $j$ via~\cite{google_qec1}:
$$p_{ij} = \frac{1}{2} - \frac{1}{2} \sqrt{1 - \frac{4(\langle x_i x_j \rangle - \langle x_i \rangle \langle x_j \rangle)}{1 - 2\langle x_i \rangle - 2\langle x_j \rangle + 4\langle x_i x_j \rangle}}$$
When a defect is present, detection probabilities can reach or exceed 0.5, causing the denominator to vanish or the radicand to turn negative (such unphysical extremes are capped at 0.5). Even without reaching extreme limits, high detection rates from defects generate strong spurious correlations. Consequently, stabilizers directly affected by a defect exhibit high correlations with all other stabilizers across the lattice, manifesting as prominent streak patterns in the correlation matrix. This background noise hinders the identification of underlying error structures and degrades decoder efficiency. In contrast, under our defect-avoidance scheme, these correlations are appreciably reduced, cleanly exposing physical error features such as time-like fault propagation. Even though super-stabilizers exhibit elevated detection rates, correlations drop substantially compared to the unmitigated case, confirming that defect avoidance restores informative syndrome signals essential for decoding, in agreement with syndrome correlation diagnostics on IBM hardware~\cite{melbourne}.

\begin{figure}[t]
\centering
\includegraphics[width=1\linewidth]{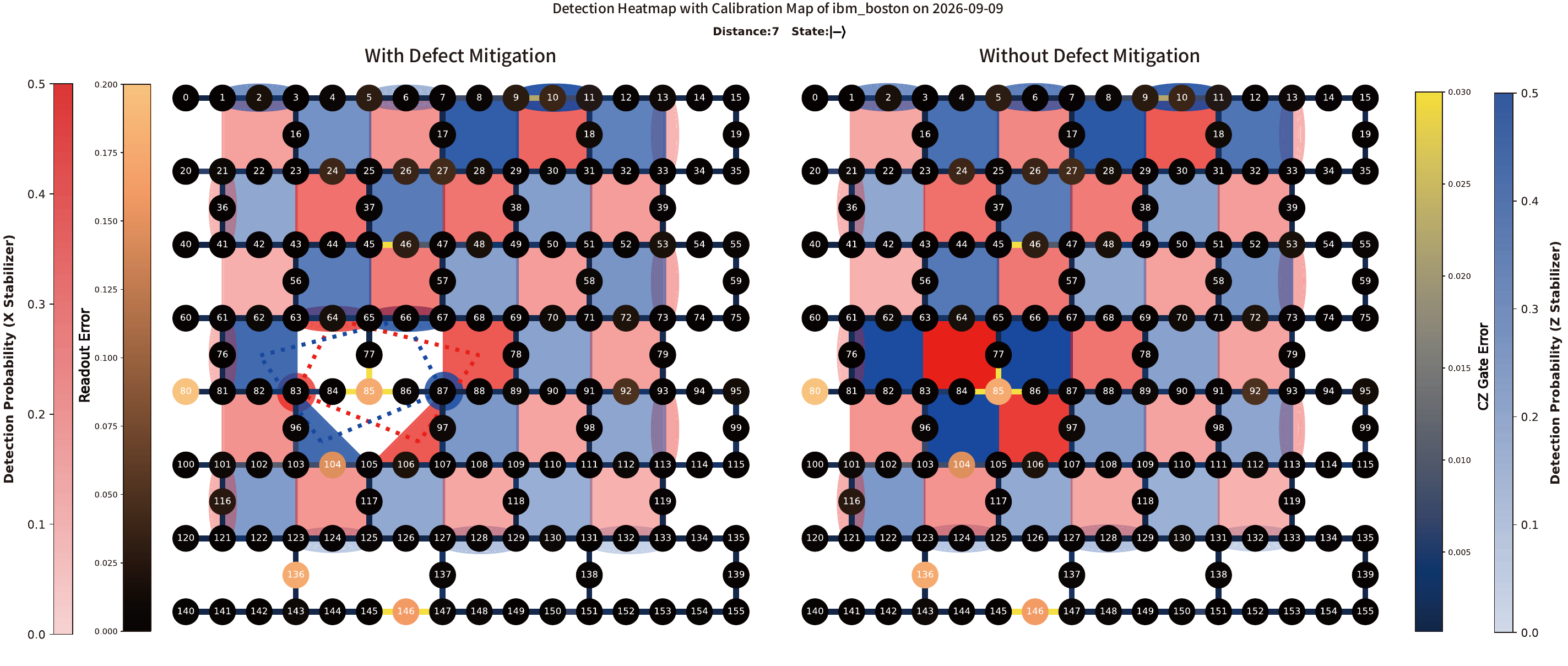} 
\caption{Detection probabilities overlaid on the calibration map of \texttt{ibm\_boston} from September 9, 2026, displaying CZ gate error rates and measurement error rates. Darker shading corresponds to higher detection probabilities. Stabilizers adjacent to high-error qubits or couplers exhibit elevated detection activity. Physical qubit~85 displays severe CZ gate and readout errors, causing all neighboring stabilizers to register high detection rates. Incorporating gauge operators into a super-stabilizer visibly reduces these localized detection events.}
\label{fig:Calibration data and detection event}
\end{figure}
We further correlated detection events directly with hardware calibration data (Fig.~\ref{fig:Calibration data and detection event}). Because implementing a distance-7 rotated surface code requires 127 of the 156 qubits on \texttt{ibm\_boston}, the resulting high qubit occupancy leaves almost no degrees of freedom to avoid defective qubits and couplers via selective code layout. In addition to primary defect qubit~85, qubits 10, 27, 46, and 104---along with their associated couplers---exhibited elevated physical error rates, resulting in increased detection probabilities on adjacent stabilizers, though less severe than qubit~85. Because detection probabilities are acutely sensitive to local physical error rates, detection maps serve as a reliable diagnostic proxy for in situ hardware performance.

The dataset presented in the main text was acquired on September 9, 2026. In an earlier run on August 18, 2026, qubit~27 also presented as a major defect with elevated detection rates. A two-defect avoidance experiment bypassing both qubits 85 and 27 was conducted on August 18, demonstrating lower logical error rates and detection probabilities than bypassing qubit~85 alone. However, following recalibration on September 9, qubit~27 improved substantially; consequently, continuing to bypass qubit~27 introduced unnecessary gauge overhead, resulting in higher logical error rates. This indicates that hardware calibration must be taken into account before selecting defects, dynamically deciding whether to deploy defect-avoidance circuits based on current device metrics (see Supplementary Information~\ref{supp:2_defects}).

\subsection{Logical error per cycle}
\begin{figure}[t]
\centering
\includegraphics[width=1\linewidth]{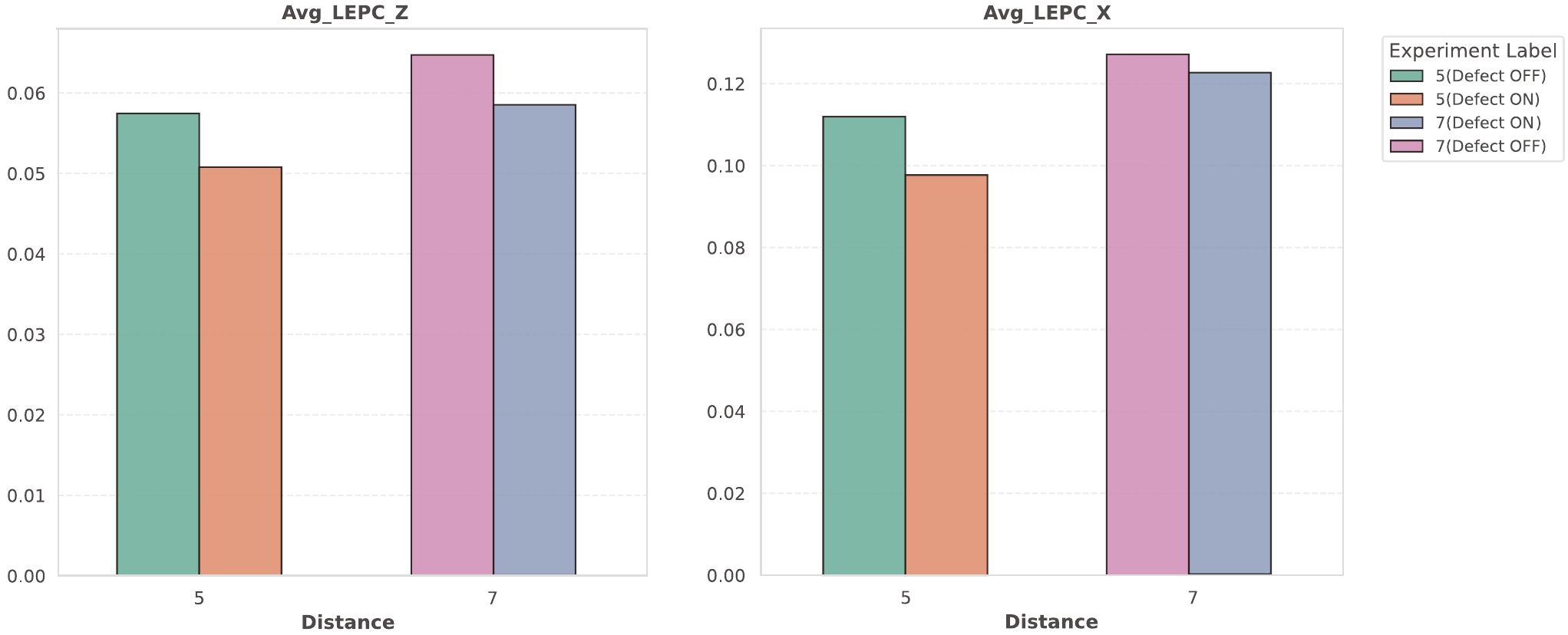} 
\caption{Comparison of logical $X$ and $Z$ error rates with and without defect mitigation for distances 5 and 7. For each distance, 10,000 experimental shots were executed for logical $X$ and $Z$ error extraction. At distance 5, the logical $Z$ error rate per cycle decreased from 0.057785 to 0.0507924 (a $\sim$12.1\% reduction), and the logical $X$ error rate decreased from 0.118848 to 0.106915 (a $\sim$10.0\% reduction). At distance 7, the logical $Z$ error rate decreased from 0.062331 to 0.057272 (a $\sim$8.1\% reduction), and the logical $X$ error rate decreased from 0.137551 to 0.128216 (a $\sim$6.7\% reduction).}
\label{fig:Defect_compare}
\end{figure}

We measured logical error rates for distances 5 and 7 while varying the number of rounds from 3 to 8 in increments of 1. Because gauge parity comparisons are temporally staggered across cycles, experiments were initiated at round 3 to accommodate this scheduling structure. For each distance, physical qubits were initialized in $|+\rangle$ ($|-\rangle$) to probe logical $Z$ errors and in $|0\rangle$ ($|1\rangle$) to probe logical $X$ errors, collecting 10,000 experimental shots per configuration with and without defect avoidance. We employed a minimum-weight perfect matching (MWPM) decoder to evaluate logical error rates (see Supplementary Information~\ref{supp:transpile_and_decoding} for details) Based on these measurements, the logical error rate per cycle (LEPC), $\epsilon_{\text{cycle}}$, was determined using:
$$
\epsilon_{\text{cycle}} = \frac{1}{2}\left(1 - (1 - 2\epsilon_L)^{\frac{1}{c}}\right)
$$
where $\epsilon_L$ denotes the cumulative logical $X$ or $Z$ error after $c$ cycles. The representative LEPC was obtained by averaging across the evaluated round durations. When defect avoidance was engaged, LEPC at distance 5 was reduced from 0.057785 to 0.0507924 for logical $Z$ (a 12.1\% reduction) and from 0.118848 to 0.106915 for logical $X$ (a 10.0\% reduction). At distance 7, the logical $Z$ error rate per cycle decreased from 0.062331 to 0.057272 (an 8.1\% reduction) and the logical $X$ error rate decreased from 0.137551 to 0.128216 (a 6.7\% reduction) (Fig.~\ref{fig:Defect_compare}). Although background physical error rates on unmitigated qubits were insufficiently low to observe net threshold error suppression with increasing distance, these results confirm significant mitigation of defect-induced logical errors. Additionally, we evaluated distance-3 rotated surface codes (see Supplementary Information~\ref{supp:distance_3}), further underscoring that hardware defects dramatically elevate logical error rates and that achieving high physical qubit fidelity is critical when scaling quantum systems.

\section{Conclusion}
In this study, we demonstrated a scalable framework for implementing rotated surface codes on heavy-hexagon architectures and introduced an efficient protocol to bypass physical defects within such constrained lattices. Experimental execution on real IBM quantum hardware verified that defect-avoidance circuits successfully suppress the disruptive effects of hardware defects, achieving an approximately 10\% reduction in logical error rates per cycle. These findings confirm that fault-tolerant architectures can be rendered scalable and resilient to defects even under low degree-2$\sim$3 connectivity. Through these results, combining defect avoidance with targeted hardware calibrations that maximize physical gate fidelity represents the most viable path toward practical fault-tolerant quantum computation.

\vspace{1em}
\noindent\textbf{Note added:}\, After completing the draft of this paper, we became aware of an independent study by McLauchlan et al.\cite{mclauchlan2026error} on defect-exclusion techniques for the dynamic compass code implemented on the same heavy-hexagon hardware.
 
\section*{Data and Code Availability}
All experimental and simulation data generated during this study are presented in the main text and supplementary information. 

\bibliographystyle{unsrt}
\bibliography{references} 

\section*{Acknowledgement}
This work is supported by the Creation of the Quantum Information Science R$\&$D Ecosystem (Grant No. 2022M3H3A1063074) through the National Research Foundation of Korea (NRF) funded by the Korean government (Ministry of Science and ICT).

\section*{Author Contributions}
H.K. conceived the core concept, designed the syndrome extraction and defect-avoidance protocols, performed the numerical simulations and experimental executions on IBM Quantum hardware, analyzed the syndrome and error data, and wrote the manuscript. Y.K. supervised the project, guided the research, and reviewed the manuscript. All authors discussed the results and contributed to the final manuscript. In accordance with journal policies, an AI language model was utilized solely for language translation assistance and grammatical refinement during manuscript preparation; all technical content, derivations, data, and scientific conclusions were formulated and verified by the authors, who take full responsibility for the work.

\section*{Competing Interests}
The authors declare no competing interests.

\section*{Supplementary information}
\setcounter{subsection}{0}
\renewcommand{\thesubsection}{\Alph{subsection}}

\subsection{Error propagation of stabilizer measurement circuit}\label{supp:error_propagation}
\begin{figure}[t]
\centering
\includegraphics[width=1\linewidth]{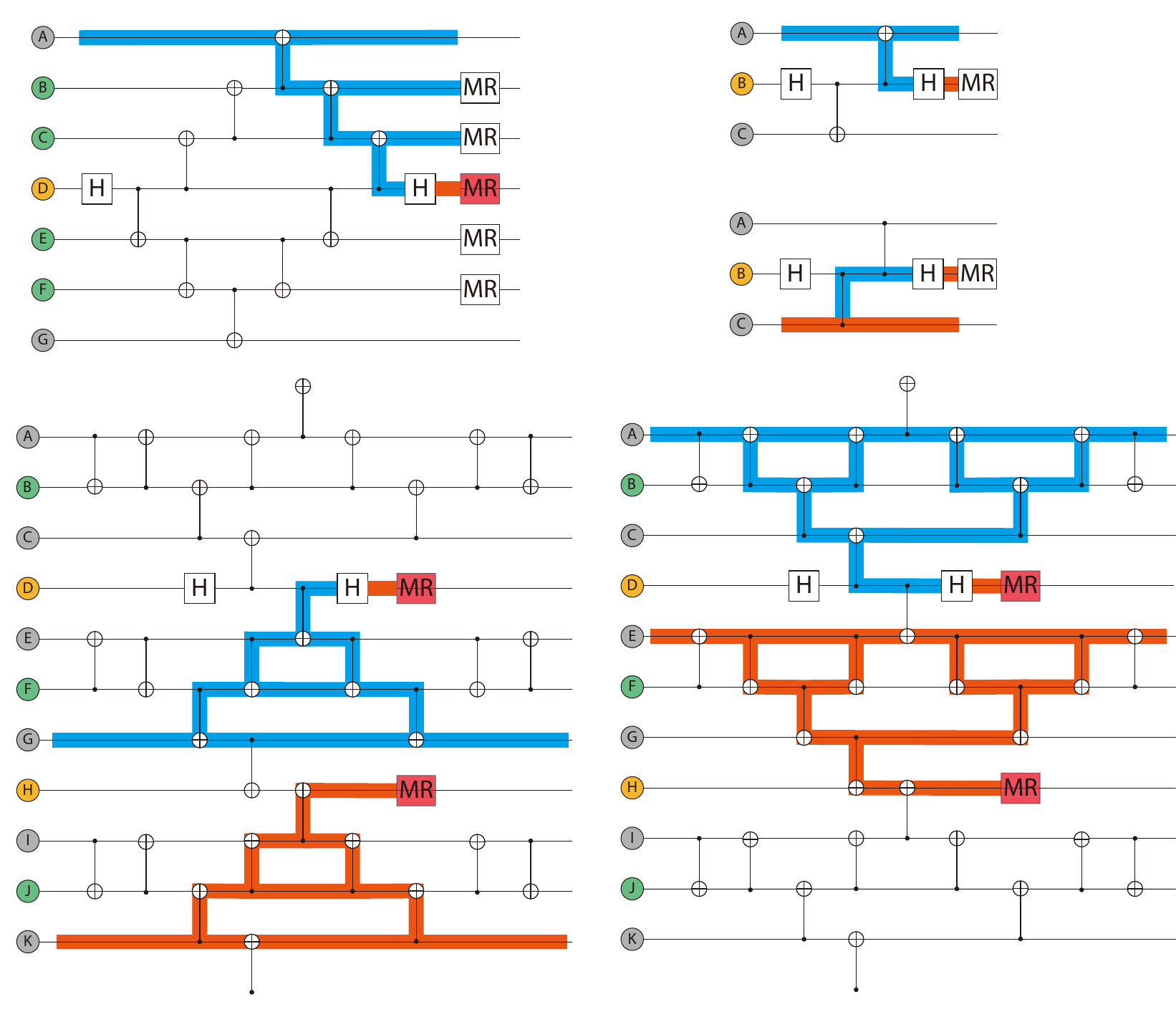} 
\caption{Error propagation diagrams for the syndrome extraction circuits shown in Fig.~\ref{fig:surface code on HH with flag}, illustrating how individual Pauli errors propagate through the gate sequences of each circuit.}
\label{fig:error propagation}
\end{figure}
Implementing stabilizer operator measurements is an essential requirement for realizing QEC codes on physical quantum hardware. The specific implementation strategy depends heavily on hardware constraints. A rigorous approach to validating the error-detecting capability of these circuits is analyzing error propagation within the stabilizer measurement schedule. Error propagation traces Pauli errors arising at arbitrary space-time locations through Clifford circuits via standard circuit identities, verifying whether the resulting error signatures trigger syndrome detection events~\cite{Fowler3, kishony2026surface}. In this study, we employed three distinct stabilizer measurement circuits. The corresponding error propagation diagrams are presented in Fig.~\ref{fig:error propagation}, confirming that all single-qubit Pauli faults on data qubits are faithfully detected. 

\begin{figure}[ht]
\centering
\includegraphics[width=0.5\linewidth]{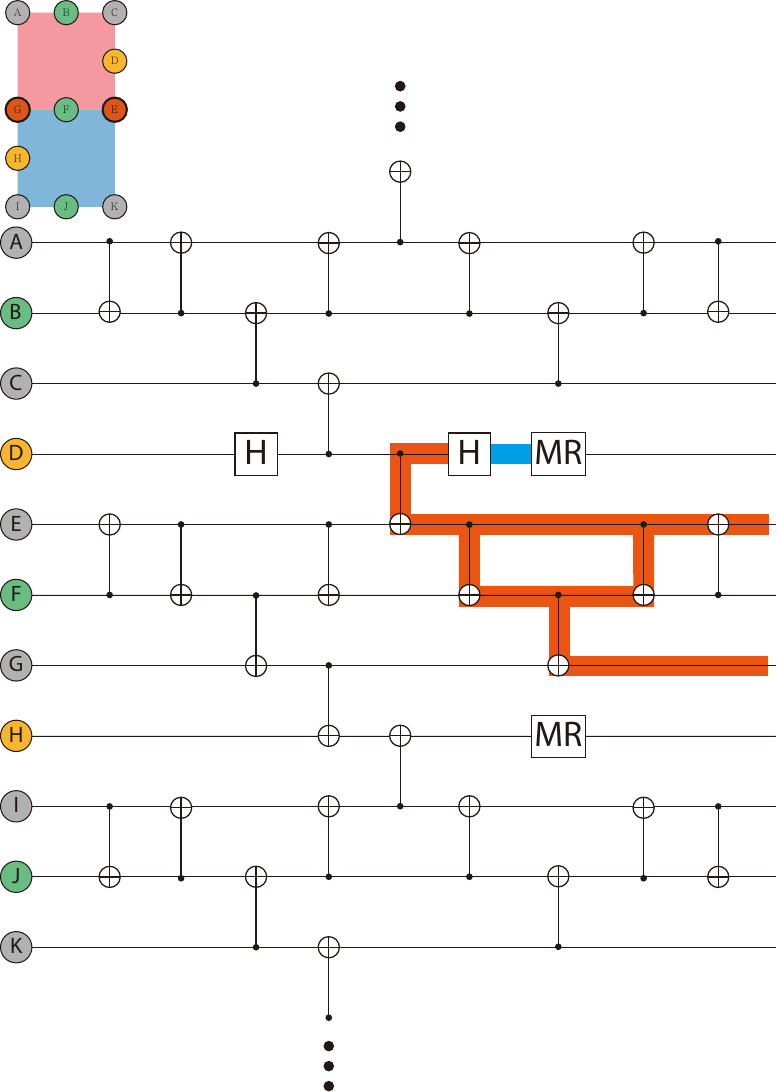}
\caption{Schematic showing an $X$ error on a syndrome qubit propagating to two data qubits. The error goes undetected at the syndrome ancilla due to the Hadamard gate, and the resulting two data qubit errors contribute as two physical faults toward a logical error.}
\label{fig:hook error}
\end{figure}

However, certain vulnerabilities exist in this design. Most notably, a single physical fault at specific circuit locations can propagate to two or more data qubits. Figure~\ref{fig:hook error} illustrates an example of such a fault mechanism. This fault escapes detection during the current stabilizer measurement round and simultaneously injects errors into two data qubits before the subsequent round. In our folded layout, these two data errors align along the horizontal logical operator, each contributing as an effective physical fault toward a logical failure. Consequently, the effective code distance against logical $X$ errors is halved, requiring only $\lceil d/2 \rceil$ faults to trigger a logical failure. This asymmetry is reflected in our experimental results, where logical $X$ error rates consistently exceed logical $Z$ error rates. Nonetheless, exponential error suppression remains feasible below the threshold (see Supplementary Information~\ref{supp:threshold}).

\subsection{Threshold Simulation of Rotated Surface Code on Heavy-Hexagon Architecture}\label{supp:threshold}

\begin{figure}[ht]
\centering
\includegraphics[width=1\linewidth]{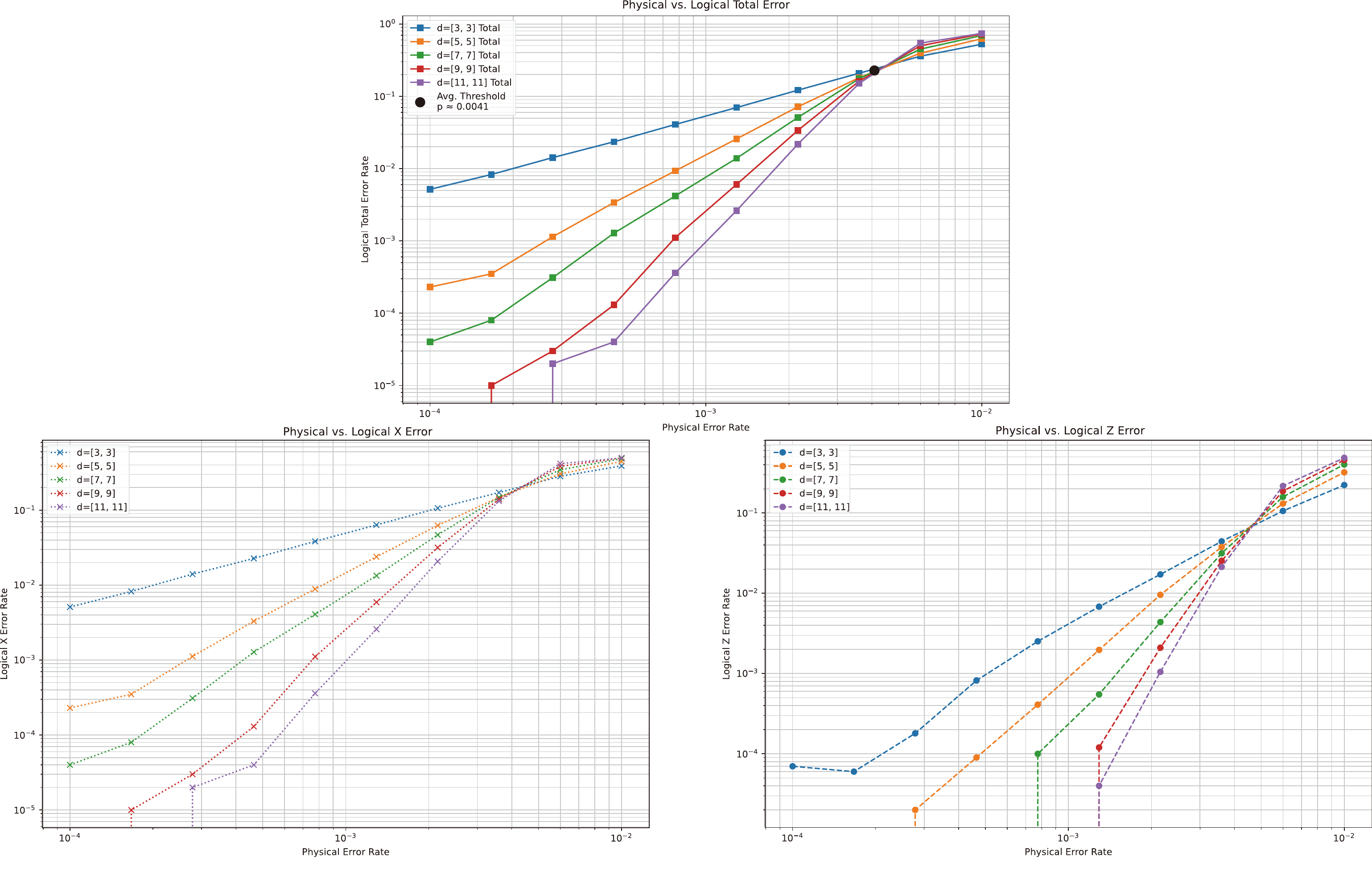} 
\caption{Simulation results of logical error rates as a function of physical error rate. Due to the reduced effective distance for logical $X$ errors, the suppression of logical $X$ errors with decreasing physical error rate is less steep than that for logical $Z$ errors. Nevertheless, despite the reduction in effective distance, a well-defined threshold is observed, demonstrating exponential error suppression below threshold.}
\label{fig:threshold}
\end{figure}

As discussed in Supplementary Information~\ref{supp:error_propagation}, applying stabilizer folding to rotated surface codes reduces the effective code distance along the folding axis. Despite this reduction, threshold simulations confirm that an error threshold below which logical error rates decrease exponentially is preserved (Fig.~\ref{fig:threshold}). Simulations were conducted in \texttt{stim}~\cite{stim}, modeling circuit-level noise with a uniform depolarizing error channel applied to each gate, following standard threshold benchmarking procedures~\cite{Fowler3}. Logical error rates were evaluated across $1 \times 10^6$ Monte Carlo samples per data point, and total logical error rates were computed from the probability that neither $X$ nor $Z$ logical failures occurred.

\subsection{Qubit Layout for Experiment}\label{supp:qubit_layout}

\begin{figure}[ht]
\centering
\includegraphics[width=1\linewidth]{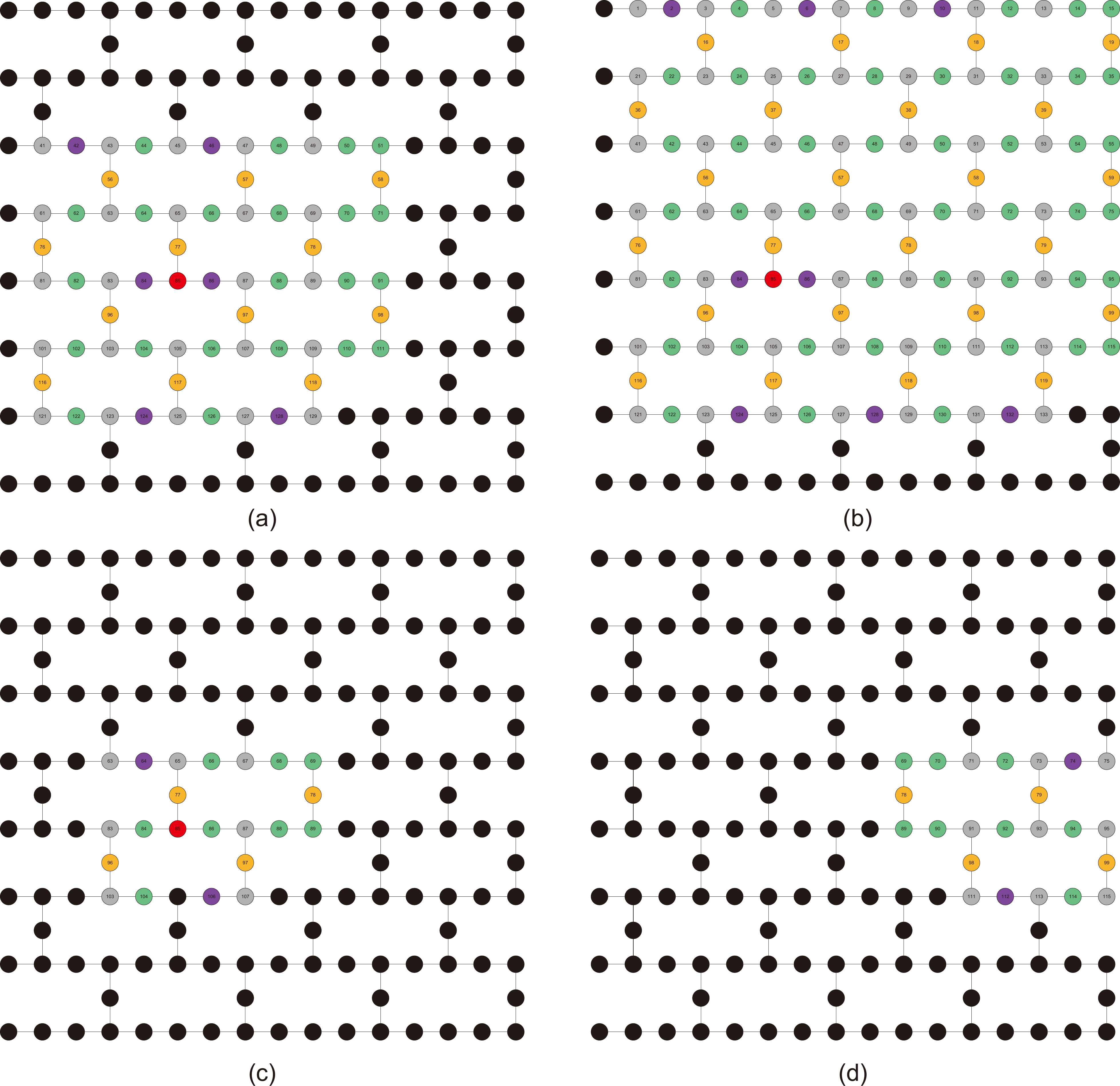} 
\caption{(a) Qubit layout used for distance 5. (b) Qubit layout used for distance 7. (c) Qubit layout used for distance 3 containing a defect. (d) Qubit layout used for distance 3 selected exclusively from low-error qubits.}
\label{fig:qubit_layout}
\end{figure}

The \texttt{ibm\_boston} processor utilized in these experiments comprises 156 physical qubits. Implementing rotated surface codes of distances 3, 5, and 7 requires 23, 65, and 127 physical qubits, respectively. Figure~\ref{fig:qubit_layout} illustrates the physical qubit layouts deployed in each experiment. The layouts for distances 5 and 7 represent configurations designed to bypass physical defect qubit~85 described in the main text, while the distance-3 layouts correspond to the comparative defect experiments presented in Supplementary Information~\ref{supp:distance_3}.

\subsection{Transpilation and Decoding}\label{supp:transpile_and_decoding}

\begin{figure}[t]
\centering
\includegraphics[width=0.8\linewidth]{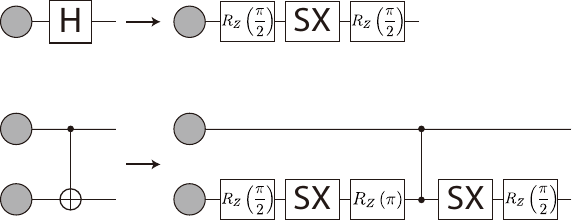} 
\caption{Decomposition of Hadamard and CNOT gates into native $R_Z$, $\sqrt{X}$ (SX), and CZ gates on IBM Heron processors.}
\label{fig:Transpile}
\end{figure}

In formulating the QEC circuits, we expressed operations in terms of Hadamard and CNOT gates. On physical quantum processors, however, these gates are not natively supported and must be decomposed into native gate sets. We synthesized Hadamard and CNOT operations using $R_Z$, $\sqrt{X}$ (SX), and CZ gates, as illustrated in Fig.~\ref{fig:Transpile}. Circuit optimization and pulse scheduling were performed using the Qiskit transpiler. To suppress decoherence during idle intervals between gates, dynamical decoupling sequences comprising four $X$ pulses were inserted~\cite{yang2011preserving}. Active qubit reset following flag and syndrome measurements was realized conditionally: a corrective $X$ pulse was applied conditioned on an outcome of 1.
The compiled circuits were executed on IBM Quantum hardware via the cloud, and resulting bitstrings were processed classically to calculate syndrome strings. To construct the detector error model for decoding, we ingested daily calibration metrics from IBM, including single-qubit gate errors (SX and $X$ gate error rates are reported identically by IBM; $R_Z$ gates are implemented as virtual phase shifts and incur zero gate error), two-qubit CZ gate errors, and readout errors. These parameters defined the edge weights in the decoding graph. Gate errors were modeled as depolarizing channels parameterized by the reported error rates, and readout errors were modeled as bit-flip ($X$) errors preceding measurement. Minimum-weight perfect matching (MWPM) implemented via PyMatching was utilized to decode the detection events~\cite{Edmonds, Kolmogorov, Fowler3, pymatching, pymatching2, ibm_qec3}.

\subsection{Gate Timeline}\label{supp:gate_timeline}

\begin{figure}[t]
\centering
\includegraphics[width=1\linewidth]{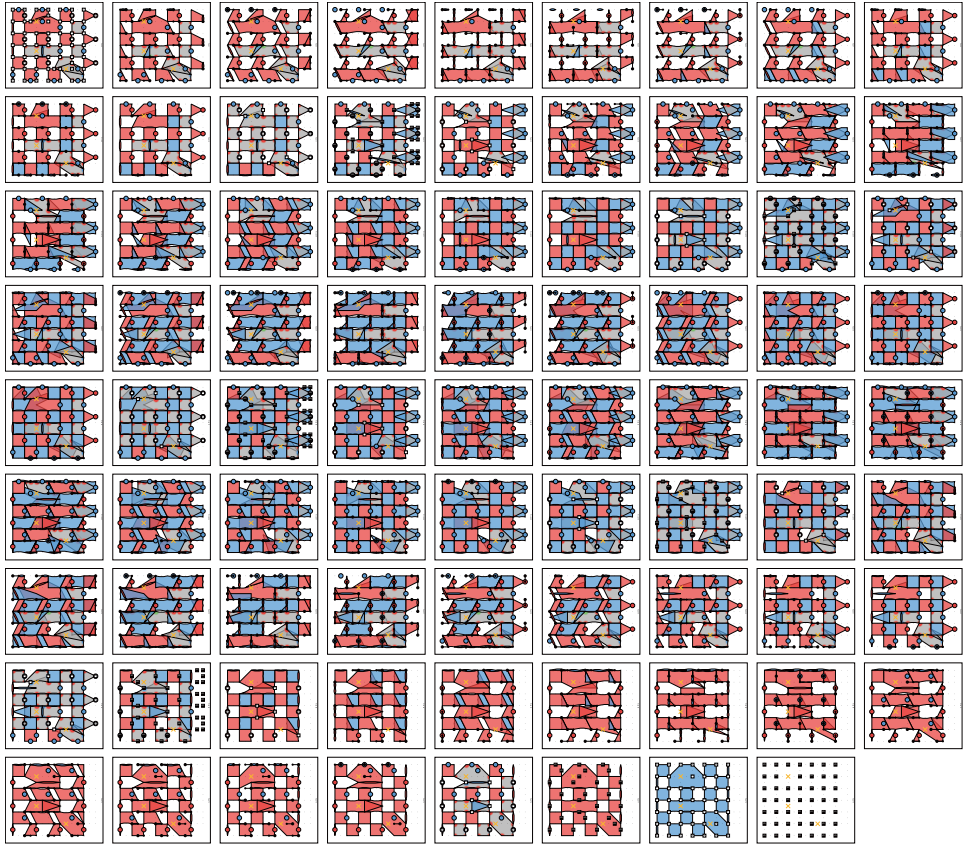} 
\caption{Temporal evolution of the modified syndrome extraction circuit designed for defect handling: weight-2 and weight-3 gauge operators are implemented using conventional circuits truncated to exclude defective components, whereas weight-1 gauge operators repurpose existing flag (bridge) qubits as syndrome qubits, facilitating full gauge measurement of all super-stabilizers within two cycles.}
\label{fig:gate_timeline}
\end{figure}

To bypass hardware defects without prolonging the syndrome extraction sequence, the temporal gate scheduling of the circuit is dynamically adapted (Fig.~\ref{fig:gate_timeline}). While weight-2 and weight-3 gauge operators are measured by truncating conventional extraction circuits to omit defective sites, measuring weight-1 gauge operators relies on repurposing existing flag (bridge) qubits as syndrome ancillas, thereby enabling the complete measurement of gauge operators for all super-stabilizers within two cycles.

\subsection{Handling Multiple Defects}\label{supp:2_defects}

\begin{figure}[t]
\centering
\includegraphics[width=1\linewidth]{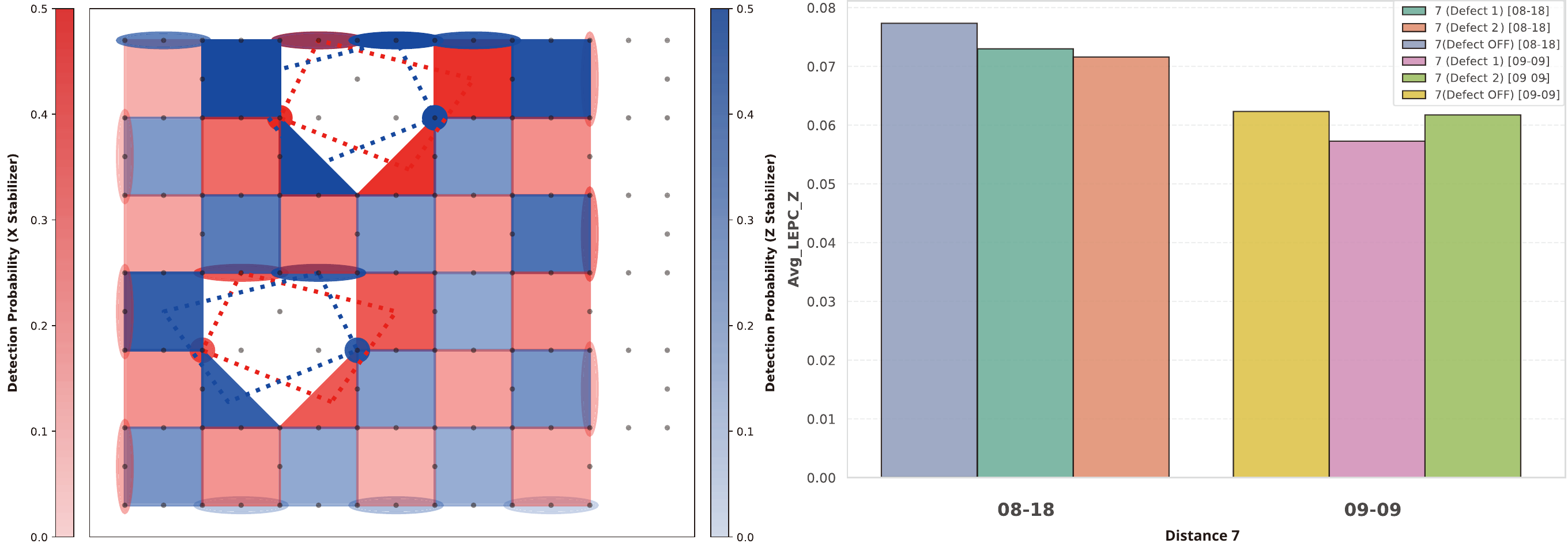} 
\caption{Left: Stabilizer configuration and detection probabilities when treating both qubits 27 and 85 as defects on August 18, 2026. Right: Variation in logical $Z$ error rate as a function of the number of bypassed defects on August 18 versus September 9. On August 18, bypassing two defective qubits yielded lower logical error rates, whereas on September 9, bypassing both qubits increased the logical error rate.}
\label{fig:2_defect}
\end{figure}

Our proposed defect-avoidance protocol can be simultaneously extended to multiple defects, provided that the respective super-stabilizers do not overlap. Furthermore, when bypassing two independent defects, the effective distance may decrease along only one basis ($X$ or $Z$) while remaining unaffected along the other, depending on their relative spatial configuration. This occurs because when neither defect lies along the shortest error chain connecting opposing boundaries, defect avoidance merely introduces an alternate equal-length path. If two defects are adjacent such that their super-stabilizers overlap, a merged super-stabilizer must be formulated; however, such overlapping configurations are beyond the scope of this study.

On August 18, 2026, we performed distance-7 experiments treating qubit~27 on \texttt{ibm\_boston} as an additional defect alongside qubit~85. The results demonstrated a lower logical error rate compared to mitigating qubit~85 alone. However, when repeating the identical experiment on September 9, recalibration of \texttt{ibm\_boston} improved the performance of qubit~27 (Fig.~\ref{fig:2_defect}). Under this improved calibration, treating qubit~27 as a defect resulted in a higher logical error rate due to the added gauge measurement overhead. These observations emphasize that hardware calibration is a critical determinant when selecting defect mitigation targets. Specifically, overall error rates were lower in the September 9 dataset, demonstrating that achieving high-quality calibration to minimize physical error rates across the processor is paramount.

\subsection{Distance-3 Rotated Surface Code Experiment}\label{supp:distance_3}

\begin{figure}[t]
\centering
\includegraphics[width=1\linewidth]{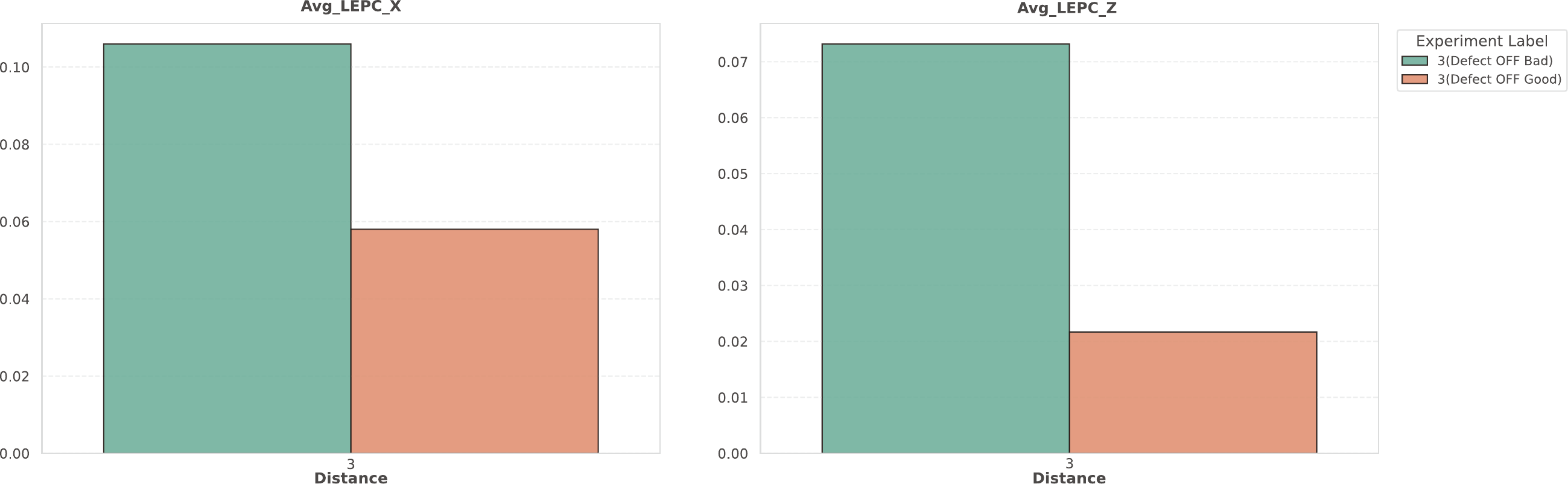} 
\caption{Comparison of logical error rate per cycle for distance 3 with and without a physical defect. In the presence of a defect, the error rate is comparable to distances 5 and 7; however, selecting an optimal sub-lattice consisting exclusively of high-fidelity physical qubits yields a substantially lower logical error rate.}
\label{fig:distance_3}
\end{figure}

In the main text, we presented experimental results for distances 5 and 7 on the heavy-hexagon lattice. Naturally, the same mapping applies to distance-3 rotated surface codes. However, because a distance-3 code has an effective distance of only 2 along the folded axis, applying defect avoidance reduces the effective distance to 1, completely forfeiting error-correcting capability. Consequently, we did not perform defect-avoidance experiments at distance 3. Instead, we leveraged distance 3 to directly quantify the impact of physical defects on logical performance. Because distance-3 codes require only 23 qubits, we can embed them within a high-fidelity sub-lattice. We compared a distance-3 code encompassing defective qubit~85 against a pristine distance-3 code placed on this optimal sub-lattice (see Fig.~\ref{fig:qubit_layout}). As shown in Fig.~\ref{fig:distance_3}, the pristine configuration achieved a markedly lower logical error rate than the defective code. Furthermore, for codes containing defective qubit~85, distance 5 exhibited a lower logical error rate than distance 3. While this might superficially resemble error suppression, physical defects dominate both cases, making it premature to attribute this solely to fault tolerance. Critically, because the pristine distance-3 code outperformed larger-distance codes plagued by defects, eliminating defects and maintaining uniform high physical fidelity is vital for realizing genuine fault-tolerant scaling.

\end{document}